\documentclass[aps,prapplied,twocolumn]{revtex4}%
\usepackage{amsfonts}
\usepackage{amsmath}
\usepackage{amssymb}
\usepackage[caption=false]{subfig}
\usepackage{graphicx}
\usepackage{verbatim}
\usepackage{color}
\usepackage{soul,xcolor}
\usepackage[normalem]{ulem}%
\providecommand{\U}[1]{\protect\rule{.1in}{.1in}}
\begin{document}
\title{Electric noise spectra of a near-surface nitrogen vacancy center diamond with
a protective layer}
\author{Philip Chrostoski$^{\ast}$}
\author{H. R. Sadeghpour$^{\dag}$}
\author{D. H. Santamore$^{\ast,\dag}$}
\affiliation{$^{\ast}$Department of Physics and Engineering Physics, Delaware State
University, Dover, DE 19901, USA}
\affiliation{$^{\dag}$ITAMP, Harvard-Smithsonian Center for Astrophysics, Cambridge,
Massachusetts 02138, USA}

\begin{abstract}
Surface noise is a detrimental issue for sensing devices based on shallow
nitrogen vacancy (NV) color center diamonds. A recent experiment indicates
that electric field noise is significant compared to magnetic field noise.
They also found that the electric field noise can be reduced with a protective
surface layer, though the mechanism of noise reduction is not well understood.
We examine the effect of a protective surface layer on the noise spectrum,
which is caused by surface charge fluctuations. We use the
fluctuation-dissipation theorem to calculate and analyze the noise spectrum
for six different surface layer materials typically used for NV center diamond
devices. We find that four parameters largely affect the noise spectrum:
effective relaxation time, effective loss tangent, power law exponent of the
noise spectrum, and layer thickness. Our results suggest that a surface
covering layer is indeed useful for decreasing surface noise, but which
material is most suitable depends on the device operational frequency range.

\end{abstract}
\date{\today }
\maketitle

%\pacs{05.40.-a, 77.22.Gm, 07.50.Hp}

\section{Introduction}

Nitrogen vacancy (NV) center diamonds are attractive candidates for a wide
range of applications, ranging from quantum metrology and sensing, to quantum
information processing, and hybrid quantum systems\cite{LRM14,ZSJ13}. Much of
the interest in NV center diamonds stems from the long quantum coherence time
of their spin states---several milliseconds at temperatures well above room
temperature\cite{BNT09}--- and their extremely high sensitivity to electric
and magnetic fields\cite{GIM08,JPJ08,MSP13,TCT10,WNH11,BSS17}. NV color
centers have a number of practical
applications\cite{AJC10,MVA14,AFC13,TPC08,CLD08,VMC09,AZP16,FHM11,DDF12,FMJ14,PKB14,ETN13,DCD13,GPN13,NMR08,NKN10,NBS10,WBS11}
in magnetic field sensing, magnetometry, scanning thermal microscopy and
frequency-modulated radio reception in extreme conditions, in addition to
being candidates for room temperature quantum computing.

When functioning as detectors, NV centers must be placed as close to the
sample surface, for maximal detection
sensitivity\cite{SSP13,RMS15,HSR15,MHM14,MHK15,FSD16}. Yet room-temperature
devices with NV centers near the surface tend to be noisy. This noise produces
fluorescent line-broadening and decreases the overall detection sensitivity.
Previously, magnetic field noise was regarded as a dominating source of
noise\cite{TAK14,BAM14,OPC12,RDS07}. The majority of magnetic field noise
comes from bulk impurity interactions of nuclear and electronic spin
baths\cite{BPB12,BPJ13}. The surface magnetic noise also exists and has been
attributed to electron spins of dangling bonds\cite{OSV09,SZZ80}, terminating
surface atoms\cite{JT09,LPM13}, absorption of molecules\cite{BVW72}, and
interactions with paramagnetic surface molecules\cite{BBK09}. Magnetic surface
states have been experimentally observed for both bulk and single crystalline
surfaces\cite{OPC12,KO13,OHB12}.

However, recent experiments by Kim \textit{et al.}\cite{MHK15} and Romach
\textit{et al.}\cite{RMT15} revealed that, in devices at room temperature,
electric field fluctuations can sometimes be a larger source for noise than
magnetic field fluctuations. The electric field noise may be caused by lattice
strain\cite{BSS17} and dipole fluctuations of an NV center due to interaction
with fluctuating surface charges. Although, the effect of the former seems to
be much smaller than that of the latter. Mamin \textit{et al.}\cite{HMM13}
were able to reduce NV center noise by placing a layer of polymethyl
methacrylate atop the diamond. Kim \textit{et al.}\cite{MHK15} also reported
noise spectrum and T$_{2}$ coherence time improvement with certain covering
materials on NV center diamonds. Other materials, however, either reduce or
increase surface noise\cite{PAN17,DDJ17,NMP09,RMT15}. Until now there has been
no systematic theoretical study of electric noise in NV center diamond devices
with different covering layers, nor any heuristics beyond experimental
intuition for determining which materials efficiently reduce noise in which
frequency bands.

In this paper, we study the physics of surface electric field noise in
diamonds containing NV centers. We evaluate the effect of noise reduction with
a liquid or solid cover layer, and identify the parameters that control noise
reduction in various situations. Since most NV center devices operate at room
temperature, we treat the surface electric field noise as being produced
mainly by thermally-activated fluctuating dipoles caused by the surface
charges. These dipole fluctuations lead to fluctuations in the electric field
at the position of the color center. We used the fluctuation-dissipation
theorem for noise calculations. Kuehn et al., for example, have used the
method to calculate dielectric fluctuation due to noncontact friction for PMMA
\cite{Kuehn06} at a fixed frequency. We will use it to calculate frequency
dependent noise with different covering materials to explore the optimal
frequencies of operation for those materials.

Our theoretical model, described in Sec.\ \ref{sec_model}, consists of a
diamond coated with a cover layer, which we also sometimes refer to as a
\textquotedblleft protective layer\textquotedblright. We use the
fluctuation-dissipation theorem to obtain noise spectra at room temperature,
and we calculate the effective capacitance and loss tangent.\ The surface
cover materials we consider are: glycerol, propylene carbonate (PC),
polymethyl methacrylate (PMMA), polyvinylindene flouride (PVDF),
perflouropolyether (PFPE), and dimethyl sulfoxide (DMSO). PMMA, PVDF, PFPE,
glycerol, and PC have been commonly used in experimental
work\cite{RMT15,HMM13,PAN17,DDJ17,NMP09}. PMMA and PVDF are both solids, while
all the others are liquid at room temperature. We analyze the noise spectra at
frequencies ranging from $1$ \textrm{kHz} $-$ $10$ \textrm{MHz}, typical in
experiments, and compare the results with experimental data of
Ref.\ \cite{MHK15}.

We find that there are four main parameters that shape the noise spectrum: the
effective relaxation time, the effective loss tangent, the coefficient of the
power law of the noise spectrum, and the thickness of the surface layer. The
effective capacitance also matters, but to a lesser extent. In
Sec.\ \ref{sec results}, we examine the role of each parameter in depth and
discuss how the protective layer affects the parameters and the shape of the
noise spectrum. The effective relaxation time $\tau_{\text{eff}}$\ determines
the transition frequency of the noise spectrum.

Our study of six experimentally representative covering materials shows that
most covering materials commonly used by experimentalists can reduce noise in
the frequency range of interest. Most liquid surface layers are much better
than solid layers at reducing noise. This is because liquid layers have
shorter (picosecond--nanosecond) characteristic relaxation times than solid
layers (microseconds). However, the power law exponent for liquids is in
general smaller than that for solids and greatly affected by the thickness.

The frequency range where power law dominates the noise spectrum is narrower
for liquids than for solids. In liquids, a constant noise floor (i.e., white
noise) reappears at frequencies above $10^{5}$--$10^{6}$ \textrm{Hz}, while
the noise spectrum in solids still follows the power law in this frequency
range. As a result, around $1$ \textrm{MHz}, solids start to perform better in
noise reduction than some liquids. At high frequencies ($>10^{7}$
\textrm{Hz}), a solid protective layer will outperform all liquid layers in
noise reduction. The thickness of the surface layer affects the amount of
noise: in general, thinner protective layers reduce white noise floor better
than that of thick protective layer.

\section{Model and calculations \label{sec_model}}

\subsection{Noise spectrum}

\begin{figure}[ptb]
\centering
\includegraphics[width=\columnwidth]{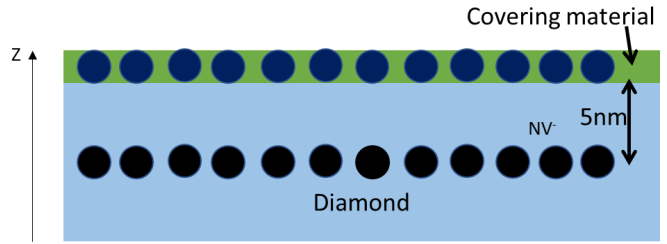}\caption{The model
consists of diamond with an NV center placed $5$ $\mathrm{nm}$ below the
surface and a protective layer covering the surface of the diamond. The lower
black dots represent the NV center, while the upper blue dots represent the
free positive charge of the protective material.}%
\label{fig1_Model}%
\end{figure}

Our model consists of a room-temperature diamond with an NV center embedded
$5$ \textrm{nm} below the diamond surface, with a liquid or solid layer
covering the surface (see Fig.\ \ref{fig1_Model}).

The covering materials we investigate are glycerol, PC, PMMA, PVDF, PFPE, and
DMSO. The reason behind choosing these different materials is due to their
favorable characteristics. For example, Kim et al.\cite{MHK15} has shown that
glycerol does not affect the dark spin density of NV centers even though
glycerol's hydroxyl groups can donate protons and possibly passivate dangling
electrons. On the other hand, propylene carbonate and DMSO are aprotic
solvents with tightly bound atoms and are optically transparent in the
wavelength regimes of NV center experiment. PMMA, PVDF, and PFPE are all
non-reactive, thermally stable, and optically
transparent\cite{Zidan03,Duan03,Jones94,Akmarov13}. PFPE has been used in
space-based applications for its optical transparency, and PMMA acts like
normal window glass only filtering out wavelengths below $300$ \textrm{nm}%
\textbf{. }

We assume that the NV centers are far apart and do not interact with each
other. We focus on the noise above the NV centers and ignore the noise below
them, since we are interested in noise reduction at the surface due to various
protective layers. The noise from below the NV centers is the same for all
covering layers, and so this noise source is not relevant for comparative
noise reduction. 

The interaction Hamiltonian of this system is%
\begin{equation}
H_{\mathrm{int}}=-\mathbf{p}\cdot\mathbf{E},
\end{equation}
where $\mathbf{p}$ is the electric dipole moment between the free positive
charges in the covering liquid and the NV center electron and $\mathbf{E}$ is
the electric field.

We want to find the noise spectrum, $S(\mathbf{k},\omega)$, arising from the
fluctuation of the dipole moment between the NV center and the covering layer.
The electric field is due to charge fluctuations and acts on the system
through the dipole moment. To obtain the noise spectrum, it s necessary to
calculate the two-point space-time correlation of the dipole moment
fluctuation, $\left\langle \delta p(\boldsymbol{r},\tau)\delta
p(\boldsymbol{r}^{{\prime}},\tau+t)\right\rangle $, and then, calculate the
noise spectrum by%
\begin{equation}
S(\mathbf{k},\omega)=2\int_{-\infty}^{\infty}\left\langle \delta
p(\boldsymbol{r},\tau)\delta p(\boldsymbol{r}^{{\prime}},\tau+t)\right\rangle
e^{-i\omega t}dt. \label{SwithPcorrelation}%
\end{equation}
However, since the system's response function is local, fluctuations at
distinct spatial coordinates are uncorrelated. Then $\delta p(\boldsymbol{r}%
,\tau)=\delta p(\tau)$ and $S(\mathbf{k},\omega)=S\left(  \omega\right)  $.
Using the fluctuation-dissipation theorem (see Appendix for details), we
obtain
\begin{equation}
S\left(  \omega\right)  =\frac{4\hbar}{1-e^{-\hbar\omega/k_{\mathrm{B}}T}%
}\varepsilon_{0}\chi_{e}^{\prime\prime}.\nonumber
\end{equation}
where $\chi_{e}^{\prime\prime}$ is the imaginary part of the complex electric
susceptibility, $k_{\mathrm{B}}$ is the Boltzmann constant, and $T$ is the temperature.

Let $\varepsilon^{\prime}$\ and $\varepsilon^{\prime\prime}$\ be the real and
imaginary parts of the permittivity. We rewrite $\varepsilon_{0}\chi
_{e}^{\prime\prime}$ in terms of the loss tangent, $\tan\phi(\omega
)\equiv\varepsilon^{\prime}/\varepsilon^{\prime\prime}$, from the complex
permittivity,
\begin{equation}
\varepsilon_{0}\chi_{e}^{\prime\prime}(\omega)=\varepsilon^{\prime}%
(\omega)\tan\phi(\omega).
\end{equation}
We choose the axis of interaction of $\mathbf{p}$ to be the $z$-axis as shown
in Fig.\ \ref{fig1_Model} and model the two layers above the NV center---the
diamond and cover layers---each with capacitance $C$. The system is analogous
to parallel-plate capacitors in series due to the intrinsic electric field
naturally generated by the dipole interactions between the negative charge in
the NV center and the positive charge in the covering layer. After some
manipulation, the noise spectrum per unit volume becomes%
\begin{equation}
\bar{S}(\omega)=\frac{4\hbar C(\omega)}{1-e^{-\frac{\hbar\omega}{k_{B}T}}}%
\tan\phi(\omega). \label{noise density}%
\end{equation}

\subsection{Effective capacitance and effective loss tangent}

Equation \ref{noise density} contains the capacitance $C(\omega)$ and loss
tangent, $\tan\phi(\omega)$. The capacitance expressed in terms of the
electric field as $C(\omega)=1/E(\omega)d$, with
\begin{equation}
E(\omega)=\frac{\kappa q}{\varepsilon^{\prime}\left(  \omega\right)  d^{2}},
\label{Efield}%
\end{equation}
where $\kappa=1/4\pi\varepsilon_{0}$, $q$ is the charge, and $d$ is the
thickness of the layer. This expression lets us write capacitance in terms of
electric permittivity. The frequency-dependent permittivity for a single
material is a relaxation permittivity given by the Havriliak-Negami relaxation
function\cite{HN67}%
\begin{equation}
\varepsilon\left(  \omega\right)  =\varepsilon_{\infty}+\frac{\triangle
\varepsilon}{\left(  1+\left(  i\omega\tau\right)  ^{\gamma}\right)  ^{\beta}%
}, \label{relaxperm}%
\end{equation}
where $\triangle\varepsilon=\varepsilon_{\infty}-\varepsilon_{s}$ is the
difference between the limiting high-frequency permittivity $\varepsilon
_{\infty}$ and the limiting low-frequency \textquotedblleft
static\textquotedblright\ permittivity $\varepsilon_{s}$, and $\tau$ is the
medium's characteristic relaxation time, given in terms of the material's
relaxation frequency $f_{r}$ by $\tau=1/2\pi f_{r}$. The exponents $\gamma$
and $\beta$ are fractional shape parameters describing the skewing and
broadening of the dielectric function.

For a diamond without any protective layer, the capacitance $C(\omega)$ and
loss tangent $\tan\phi(\omega)$ in Eq.\ (\ref{noise density}) are simply
calculated from the properties of diamond. However, with a protective layer on
the surface, we need to compute and use an effective capacitance
$C_{\text{\textrm{eff}}}(\omega)$\ and loss tangent $\tan\phi
_{\text{\textrm{eff}}}\left(  \omega\right)  $ to calculate $\bar{S}(\omega)$.
We model the structure as a \textquotedblleft covering layer\textquotedblright%
\ above the top diamond layer \textit{(i.e.}, the $5$ \textrm{nm} diamond
layer directly above the NV center). Then, the effective loss tangent
$\tan\phi_{\text{eff}}\left(  \omega\right)  $ is given by
\cite{MIV15,VBR10,HGS02,HAB06},%
\begin{equation}
\tan\phi_{\text{eff}}\left(  \omega\right)  =b_{H}d_{H}\tan\phi_{H}%
(\omega)+b_{L}d_{L}\tan\phi_{L}\left(  \omega\right)  , \label{TwoLayerLoss}%
\end{equation}
where $H$ and $L$ respectively denote the higher and lower index of refraction
of the surface liquid and the top diamond layer. The coefficients $d_{L}$ and
$d_{H}$ are the two materials' thicknesses, and%
\begin{equation}
b_{L,H}=\frac{1}{\sqrt{\pi}w}\left(  \frac{Y_{L,H}}{Y_{s}}+\frac{Y_{s}%
}{Y_{L,H}}\right)  , \label{coeff_b}%
\end{equation}
where $w$ is the width and $Y_{L,H,s}$ is the Young's modulus. The subscript
$s$ refers to the substrate, that is, the diamond below the NV center. We have
written this formula as $Y_{L,H,s}$ even though $s=L$\ or $s=H$---either the
$L$ layer or the $H$ layer is diamond, as is the substrate---so that
Eq.\ (\ref{coeff_b}) can be used for two different covering layers on the
diamond surface if needed. The effective capacitance $C_{\text{\textrm{eff}}%
}(\omega)$ is the capacitance of the two layers of dielectric materials,
equivalently to the diamond and the protective layer in series. The material
parameters used in our calculations are taken from the literature\cite{ANB14}
and are shown in Table \ref{table_1}.

\begin{table}[h]
\begin{center}%
\begin{tabular}
[c]{|l|l|l|l|l|l|l|}\hline
material & $\varepsilon_{s}$ & $\varepsilon_{\infty}$ & Y$\left(
\mathrm{GPa}\right)  $ & $\tau\left(  \mathrm{ns}\right)  $ & $\gamma$ &
$\beta$\\\hline
diamond & 7.99 & 5.7 & 443 & 2480 & 0.97 & 0.89\\\hline
glycerin & 51.8 & 3.9 & 4.80 & 19.9 & 1.00 & 0.67\\\hline
PC & 64.9 & 4.7 & 8.09 & 0.0433 & 0.985 & 0.927\\\hline
DMSO & 45.9 & 6.36 & 4.50 & 11.0 & 1.00 & 1.00\\\hline
PMMA & 3.60 & 2.6 & 5.00 & 500 & 1.00 & 1.00\\\hline
PVDF & 7.50 & 5.0 & 1.10 & 100000 & 1.00 & 1.00\\\hline
PFPE & 104 & 5.0 & 34.5 & 7.00 & 0.87 & 1.00\\\hline
\end{tabular}
\end{center}
\caption{Material parameters used in Eq.\ (\ref{coeff_b}) and
Eq.\ (\ref{relaxperm})}%
\label{table_1}%
\end{table}

\section{Results and discussion\label{sec results}}

\begin{figure}[ptb]
\centering
\includegraphics[width=\columnwidth]{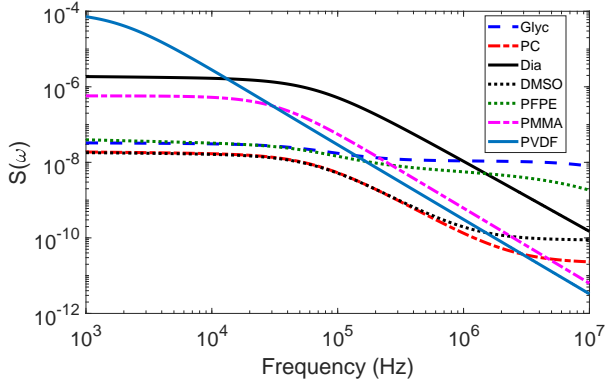}\caption{Charge noise
spectrum: bare diamond (black); with glycerol (blue dashed); propylene
carbonate (red dashed); dimethyl sulfoxide (black dotted); perfluoropolyether
(green dotted); polyvinylidene fluoride (navy); and polymethyl methacrylate
(pink dashed). The NV center diamond with a liquid protective layer exhibits
better noise reduction than the bare diamond and the diamond with a solid
protective layer in the frequency range $<10^{6}$ \textrm{Hz}.}%
\label{fig2_Noise}%
\end{figure}

Fig.\ \ref{fig2_Noise} shows calculated noise spectra as a function of
frequency for bare diamonds and for diamonds with various surface protective
layers of thickness $5$ $\mathrm{nm}$.

At low frequencies the noise spectra exhibit white noise, with a transition at
higher frequencies to a power law spectrum, with $S\left(  \omega\right)
\propto1/f^{a}$. This agrees with the experimental findings of Romach
\textit{et al.}\cite{RMT15}. White noise at low frequencies comes about when
dipoles try to align themselves into an equilibrium state.

We see from Fig.\ \ref{fig2_Noise} that all surface covering materials we
examined reduce the noise in some though not all frequency ranges. This agrees
with the experimental observations reported for glycerin and propylene
carbonate by Kim, \textit{et al.} \cite{MHK15}. However, certain materials are
more effective in reducing noise at low frequencies, while others work better
at high frequencies. For example, PVDF generates more noise than bare diamond
at frequencies less than$\ 3\times10^{4}$ \textrm{Hz} but surpasses PMMA in
noise reduction above $2\times10^{5}$~\textrm{Hz}, glycerol above
$4\times10^{5}$~\textrm{Hz}, and PFPE\ above $10^{6}$~\textrm{Hz}.

Four parameters largely affect the shape of noise spectrum at room
temperature: the effective relaxation time $\tau_{\text{\textrm{eff}}}$, the
effective loss tangent $\tan\phi_{\text{\textrm{eff}}}\left(  \omega\right)
$, the power law coefficient $a$, and the thickness of the surface layer. The
effective capacitance also affects the noise spectrum; however, the effect is
minimal compared to the other parameters, because of very little variation of
the real permittivity $\epsilon^{\prime}$\ with frequency (see
Sec.\ \ref{loss_tangent}).

\subsection{Effective relaxation time}

The effective relaxation time $\tau_{\text{\textrm{eff}}}$\ determines the
transition frequency $f_{t}$ at which the spectrum transitions from white
noise to a power law. The relationship is $f_{t}=2\pi/\tau_{\text{\textrm{eff}%
}}$, where $1/\tau_{\text{\textrm{eff}}}=1/\tau_{\text{\textrm{dia}}}%
+1/\tau_{\text{\textrm{surf}}}$. Here $\tau_{\text{\textrm{dia}}}$ and
$\tau_{\text{\textrm{surf}}}$ are the material relaxation times of the diamond
and the protective surface layer, respectively. Thus, for a liquid surface
protective layer, in which $\tau_{\text{\textrm{surf}}}\gg\tau
_{\text{\textrm{dia}}}$, the effective transition time is dominantly
determined by the relaxation time of the diamond. With a solid surface, on the
other hand, $\tau_{\text{\textrm{surf}}}$ is comparable to $\tau
_{\text{\textrm{dia}}}$, and thus the effective relaxation time will be
determined by both $\tau_{\text{\textrm{surf}}}$ and $\tau_{\text{\textrm{dia}%
}}$. We discuss the effect of thickness in Sec.\ \ref{sebsec:thickenss}.

\subsection{Effective loss tangent \label{loss_tangent}}

\begin{figure}[ptb]
\centering
\par
\subfloat[Imaginary
permittivity.]{
	\label{subfig:ImPerm}
	\includegraphics[width=0.5\textwidth]{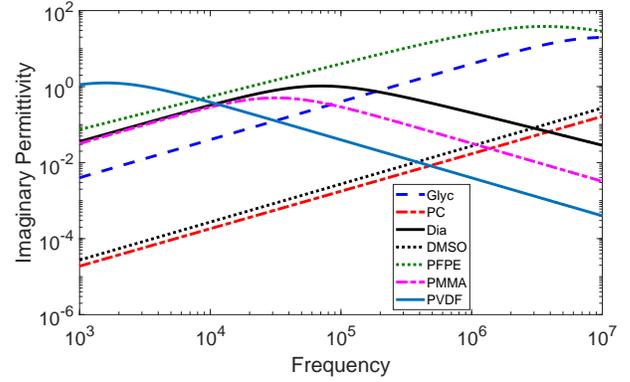} }
\par
\subfloat[Real permittivity. The real part of the permittivity is almost constant with frequency except at frequencies around $1/\tau$.]{
	\label{subfig:RePerm}
	\includegraphics[width=0.5\textwidth]{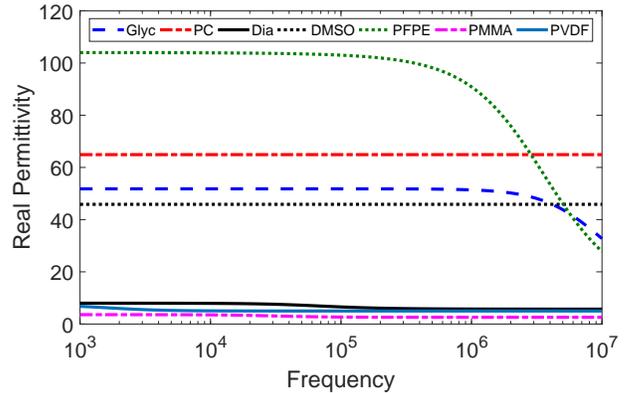} }\caption{Imaginary
and real permittivities as a function of frequency: bare NV center diamond
(black); with glycerol (blue dashed); propylene carbonate (red dashed);
dimethyl sulfoxide (black dotted); perfluoropolyether (green dotted);
polyvinylidene fluoride (navy); and polymethyl methacrylate (pink dashed)}%
\label{fig3_permittivity}%
\end{figure}

Figures\ \ref{fig2_Noise} and \ref{fig3_permittivity}(a) show that the amount
of white noise is determined by the effective loss tangent $\tan
\phi_{\text{\textrm{eff}}}\left(  \omega\right)  $. In addition, overall
charge fluctuation noise is proportional to the dielectric loss tangent, and
thus charge fluctuation noise increases for lossy materials.

Next, since loss tangent is defined as the ratio of the imaginary permittivity
$\epsilon^{\prime\prime}$ to the real permittivity $\epsilon^{\prime}$, we
look into both $\epsilon^{\prime}$ and $\epsilon^{\prime\prime}$. Compared to
the imaginary part (see Fig.\ \ref{fig3_permittivity}(a)), the real part of
the permittivity is relatively constant with frequency throughout the
frequency range $10^{3}$--$10^{7}$~\textrm{Hz}, except around the transition
frequency $f_{t}$ (see Fig.\ \ref{fig3_permittivity}(b)). As a result,
$\tan\phi_{\text{\textrm{eff}}}\left(  \omega\right)  $ is predominantly
determined by the imaginary permittivity. This could be why Kim \textit{et
al.}\cite{MHK15}, who used only real permittivity to estimate the noise, could
not explain the noise spectra they observed in their experiments.

The noise spectrum is also affected by the capacitance, as shown in
Eq.\ (\ref{noise density}), and thus it is also proportional to the real part
of the permittivity. However, since the real permittivity is relatively
constant with frequency, except around the transition frequency $f_{t}$, the
effect of capacitance on the noise spectrum is much less than that of the loss tangent.

\subsection{Power law exponent}

In atomic systems such as an ion trap, the power law $1/f^{a}$ is associated
with the number of excitation modes of environmental phonons \cite{SRWS11}.
However, in our study, temperature is held constant, and thus the excitation
modes of environmental phonons are the same for all materials. As can be read
off Fig.\ \ref{fig2_Noise}, the noise spectra of the solid systems we
considered (bare diamond, PVDF$+$diamond, and PMMA$+$diamond) follow a power
law with exponent ranging from $a=-2.1$ to $a=-2$. On the other hand, with a
liquid surface layer such as glycerin, PC, DMSO, and PFPE, the power law
exponent varies from $a=-0.5$ to $a=-1.9$. The exponent depends on thickness,
as discussed in the next subsection \ref{sebsec:thickenss}.

\subsection{Thickness of the surface layer\label{sebsec:thickenss}}

\begin{table}[h]
\begin{center}%
\begin{tabular}
[c]{|l|l|l|}\hline
material & $a\left(  d=5\mathrm{nm}\right)  $ & $a\left(  d=1\mathrm{\mu
m}\right)  $\\\hline
PC & -1.6 & -0.91\\\hline
glycerin & -0.5 & -0.15\\\hline
Silicon & -1.89 & -1.92\\\hline
DMSO & -1.9 & -1.1\\\hline
PFPE & -1.6 & -0.2\\\hline
PMMA & -2.2 & -2.1\\\hline
PVDF & -2.0 & -1.94\\\hline
\end{tabular}
\end{center}
\caption{Noise power exponent of seven materials examined with thickness
$1\mathrm{nm}$ and $1\mathrm{\mu m}$.}%
\label{table_2}%
\end{table}

Table \ref{table_2} shows that for liquid surface layers, the power law
exponent $a$ varies drastically with thickness. For solid surface layers,
though, $a$ does not vary much with thickness. The power law region occurs
when the dipole moments of the surface covering layer have aligned themselves
to the NV center. Thus, the power law of a liquid covered diamond seems to be
strongly affected by the stability of aligned dipole moments than that of a
solid covered diamond.

\begin{figure}[ptb]
\centering
\includegraphics[width=\columnwidth]{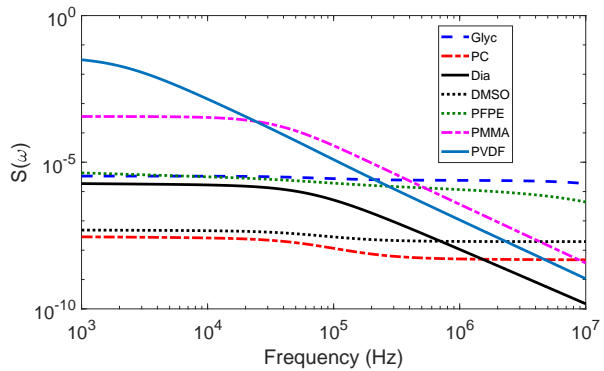}\caption{Noise
spectrum with thickness $1~\mathrm{\mu m}$. Thicker surface layers have higher
noise floor. The power law exponent changes with thickness for liquid surface
layers but almost the same for solid surface layers.}%
\label{fig5_NoiseThick}%
\end{figure}

Figure \ref{fig5_NoiseThick} shows the noise spectra with thick ($1$
$\mathrm{\mu m}$) surface protective layers. The thickness of the protective
layer does not change the noise spectrum transition frequency $f_{t}$, since
the transition is governed by the relaxation time. However, a thick surface
protective layer does increase the noise floor in general.

The exception to this rule is for those materials with a very short relaxation
time $\tau_{\text{\textrm{surf}}}$, such as PC and DMSO and only at low
frequency range (noise floor is high for high frequency range). This puzzling
phenomenon can be explained when we re-express $\bar{S}(\omega)$\ in
Eq.\ (\ref{noise density}) by explicitly writing the effective capacitance in
terms of geometry and factoring out thickness independent terms. After some
simplification, we obtain%
\begin{equation}
\bar{S}(\omega)\approx\Gamma\left(  \frac{b_{L}}{\varepsilon_{H}^{^{\prime}}%
}\frac{d_{L}}{d_{H}}\phi_{L}+\frac{b_{H}}{\varepsilon_{H}^{^{\prime}}}\phi
_{H}\right)  . \label{ThickDepenNoise}%
\end{equation}
where $\Gamma$ incorporates all factors independent of layer thickness. The
first term of Eq.\ (\ref{ThickDepenNoise}) contains the ratio of the
thicknesses of the protective layer and top diamond layer; it is the only
thickness dependent part of the equation. Note, however, that this first term
also depends on the loss tangent. As a result, in materials such as PC and
DMSO whose loss tangent $\phi_{L}$ is much smaller at low frequencies than
that of diamond, the first term will be negligible compared to the second
term. Thus, in these materials, thickness makes almost no difference to the
noise spectrum at low frequencies as seen in Fig.\ \ref{fig5_NoiseThick}.

\subsection{Which cover layers reduce noise most effectively?}

Having highlighted the roles of four parameters---relaxation time, loss
tangent, power law, and thickness---in noise reduction, we now draw out some
practical implications of our results. At low frequencies, fluctuations in
dipole-dipole interactions manifests as white noise. Because its
characteristic relaxation time is shorter than that of diamond, a thin ($<10$
$\mathrm{nm}$) protective layer can reduce noise by decreasing the effective
loss tangent. In the frequency range understudy, liquid surface layers reduce
noise more effectively than solid surface layers because the
picosecond-to-nanosecond relaxation times of liquids is much shorter than the
microsecond relaxation times of solids. Because they can move more easily than
in solids, the dipoles in liquids realign more quickly than the dipoles in
solids. The low mobility in solids and the consequently slower realignment of
dipoles leads to more fluctuation noise. On the other hand, the power law
coefficient in liquids is generally smaller than in solids, and the frequency
range over which the power law holds is narrower in liquids than in solids. In
liquids, the noise spectrum becomes flat again at frequencies beyond a few
megahertz ($10^{6}$ \textrm{Hz}), whereas in solids the power law persists
even at higher frequencies. As a result, at high frequencies ($>10^{7}$
\textrm{Hz}), solid protective layers outperform liquid protective layer in
noise reduction. Furthermore, the thickness of the surface layer also affects
the amount of noise: the thinner the protective layer, the better the noise
reduction. A solid protective surface thicker than $10$ \textrm{nm} may
increase noise rather than reduce it.

\section{Conclusions}

We have studied the effectiveness at reducing surface noise of protecting
layers on NV center diamonds. We have assumed that the main source of the
noise is thermally activated surface charge fluctuations. This assumption is
reasonable, since the NV center based devices in question operate at room
temperature, and various experiments indicate that the noise mainly originates
from the electric field\cite{RMT15,MHK15}. We have used the quantum
fluctuation-dissipation theorem to calculate the noise spectrum. We have
analyzed the noise spectra for six materials commonly used to cover the
surface of NV center diamonds: glycerol, propylene carbonate, polymethyl
methacrylate, polyvinylindene flouride, perflouropolyether, and dimethyl
sulfoxide. The covering materials examined in this work all exhibit optical
transparency in NV center experiments and do not interfere with NV center
readout\cite{Zidan03,Duan03,Jones94,Akmarov13}. While chemical reactivity with
biomolecules for sensing could possibly be affected by the coating, reduction
in noise may more than compensate for this effect.

Our results show that four parameters affect the noise spectra: the effective
relaxation time, the effective loss tangent, the exponent of the power law,
and the thickness of the surface layer. (The effective capacitance also
influences the overall noise spectra, but to a much smaller extent.) Of these
four parameters, the effective relaxation time, the effective loss tangent,
and the coefficient of the power law determine the shape of the noise spectra,
while the thickness of the surface layer determines the overall amount of
noise. The effective relaxation time $\tau_{\text{\textrm{eff}}}$\ determines
the transition frequency of the noise spectrum: the longer the relaxation
time, the lower the transition frequency. The power law behavior is associated
with the stability of aligned dipole moments. Thus solid cover materials have
higher power law exponents (around $-2$) that are very little influenced by
the thickness, while liquid cover materials exhibit a drastic decrease in the
power law coefficient as the thickness increases. Consequently, one can
control the power law behavior by adjusting the cover layer thickness, if
liquid covering layer is used. The amount of white noise is determined by a
combination of the effective loss tangent and thickness. In addition, overall
charge fluctuation noise is proportional to the loss tangent, and thus the
surface noise is worse for so-called \textquotedblleft lossy
materials\textquotedblright. In general, the noise floor considerably
increases with increasing thickness. The only exception to this rule is a
material with very short relaxation time, such as PC and DMSO, and only at low
frequency ranges. The thickness of the surface layer does not affect the
transition frequency from the white noise to a power law of the noise spectrum.

Our study of six experimentally representative covering materials shows that
most covering materials commonly used by experimentalists can reduce noise in
the frequency range of $10^{3}$--$10^{7}$ \textrm{Hz}. Liquid surface layers
reduce noise better overall than solid layers, because their relaxation times
are shorter. The ideal material for reducing NV center surface noise would be
a thin liquid layer with a low real permittivity and a fast (picosecond range)
relaxation time in the frequency range $10^{3}$--$10^{7}$ \textrm{Hz}. Solid
materials in general are noisier, but their high power law exponents
compensates for the substantial white noise and provides better noise
reduction at high frequencies ($>10^{7}$~\textrm{Hz}).

It should be noted that PMMA is shown to enhance emission that might result in
reduction in lifetime\cite{KCR15}. However, due to the complexity of the
nano-diamond tip structure, such measurements may not be directly applicable
to our case: in our case, the NV center is embedded in the bulk diamond.

Finally, the electronic field noise will decrease $T_{2}$ coherence time just
as magnetic field noise does. Since some experiments such as\cite{BSS17,MHK15}
show that the electric field noise and magnetic field noise are comparable,
reducing the electric field noise might extend $T_{2}$ time considerably.

\begin{acknowledgments}
We thank Pauli Kehayias and others in Ron Walsworth group at Harvard
University for useful discussions and providing experimental insights. DHS is
grateful to ITAMP for hosting her visit during the summer and providing a
stimulating research environment. This work is supported by the NSF grant DMR-1505641.
\end{acknowledgments}

\appendix

\section{Noise spectrum derivation\label{AppdxF-T}}

In this appendix, we derive the noise spectrum via the fluctuation-dissipation
theorem. We treat the the dipole-dipole interaction Hamiltonian between liquid
and the NV center as perturbation%
\begin{equation}
\delta H=-\mathbf{p}\cdot\mathbf{E}%
\end{equation}
where $\mathbf{p}$ is the electric dipole moment between the charges in the
covering liquid and the NV center, and $\mathbf{E}$ is the electric field. We
treat the interaction using the electric dipole approximation. At thermal
equilibrium, the ensemble average of the dipole moment is defined as,%

\begin{equation}
\left\langle \mathbf{p}\right\rangle \equiv\left\langle \mathbf{p}\left(
\mathbf{r},t\right)  \right\rangle =\frac{%
%TCIMACRO{\dint }%
%BeginExpansion
{\displaystyle\int}
%EndExpansion
f_{\mathrm{eq}}\left(  \mathbf{r}\right)  \mathbf{p}\left(  \mathbf{r}%
,t\right)  dr}{%
%TCIMACRO{\dint }%
%BeginExpansion
{\displaystyle\int}
%EndExpansion
f_{\mathrm{eq}}\left(  \mathbf{r}\right)  dr} \label{EnsAve}%
\end{equation}
where $f_{\mathrm{eq}}$ is the distribution function%

\begin{equation}
f\left(  x\right)  =f_{\mathrm{eq}}e^{-\left(  \delta H/k_{\mathrm{B}%
}T\right)  }. \label{DistFunc}%
\end{equation}
where $k_{\mathrm{B}}$ is the Boltzmann constant and $T$ is the temperature,
and the integration runs over all coordinates because the system is at
equilibrium and $t$ is time. The fluctuation is regarded as a system that has
been perturbed by $\delta\mathbf{p}$%

\begin{equation}
\delta\mathbf{p}=\mathbf{p}-\left\langle \mathbf{p}\right\rangle .
\end{equation}
This perturbation is small and linearly dependent on $\mathbf{E}$. This
dependence can be given by linear response function
theory\cite{Apra75,HJMG02,JWS84},%

\begin{equation}
\delta p_{j}\left(  t\right)  =\frac{1}{2\pi}\sum_{k}%
%TCIMACRO{\dint \limits_{-\infty}^{t}}%
%BeginExpansion
{\displaystyle\int\limits_{-\infty}^{t}}
%EndExpansion
\tilde{\alpha}_{jk}\left(  t-t^{\prime}\right)  E_{k}\left(  t^{\prime
}\right)  dt^{\prime}. \label{LinResponse}%
\end{equation}
Here $\tilde{\alpha}_{jk}$ is the response function of the system and
$j,k=x,y,z$. We have assumed that the system is stationary (\textit{i.e.}, the
response function is local) and the dependence on $t-t^{\prime}$ enforces
causality. The \textquotedblleft memory\textquotedblright\ of the system is
contained in $\tilde{\alpha}_{jk}$ and we need to determine $\tilde{\alpha
}_{jk}$.

Time-dependence of the electric field perturbation is a step function to
ensure complete relaxation of the system at times $t=0$, and the perturbation
occurs $\left[  0,\infty\right)  $ with $E=E_{\mathbf{k}}^{0}$ at $t>0$.
Evaluating Eq.\ (\ref{LinResponse}) for the step function perturbation we get,%

\begin{equation}
\delta p_{j}\left(  t\right)  =\frac{E_{k}^{0}}{2\pi}%
%TCIMACRO{\dint \limits_{t}^{\infty}}%
%BeginExpansion
{\displaystyle\int\limits_{t}^{\infty}}
%EndExpansion
\tilde{\alpha}_{jk}\left(  \tau\right)  d\tau,
\end{equation}
where $\tau\equiv t-t^{\prime}$. Solving for $\tilde{\alpha}_{jk}$ we obtain,%

\begin{equation}
\tilde{\alpha}_{jk}\left(  t\right)  =-\frac{2\pi}{E_{k}^{0}}\Theta(t)\frac
{d}{dt}\delta p_{j}\left(  t\right)  . \label{ResponsFunct}%
\end{equation}
We assumed that $\tilde{\alpha}_{jk}\left(  t\right)  $ and its
time-derivative tends to zero as $t\rightarrow\infty$ and $\Theta(t)$ is the
Heaviside step function to ensure causality. Since the ensemble average is
independent of time,%
\begin{equation}
\tilde{\alpha}_{jk}\left(  t\right)  =-\frac{2\pi}{E_{k}^{0}}\Theta(t).
\end{equation}
At time $t=0$, the system is in thermal equilibrium. Expanding the exponential
in a series and plugging into Eq.\ (\ref{EnsAve}), and keeping only the terms
up to linear order in $\delta H$ we obtain,%

\begin{equation}
\mathbf{p}=\left\langle \mathbf{p}\right\rangle -\frac{1}{k_{B}T}\left[
\left\langle \delta H\left(  s\right)  \mathbf{p}\left(  s,t\right)
\right\rangle -\left\langle \mathbf{p}\left(  s,t\right)  \right\rangle
\left\langle \delta H\left(  s\right)  \right\rangle \right]  .
\label{EnsAveExpand}%
\end{equation}
Since $\delta H\left(  s\right)  $ is the perturbation at time $t=0$ we have
$\delta H\left(  s\right)  =-\mathbf{p}\left(  s,0\right)  E_{\mathbf{k}}^{0}$
and Eq.\ (\ref{EnsAveExpand}) becomes,%

\begin{equation}
\delta p\left(  t\right)  =\frac{E_{\mathbf{k}}^{0}}{k_{B}T}\left\langle
\delta p_{k}\left(  0\right)  \delta p_{j}\left(  t\right)  \right\rangle .
\end{equation}
Inserting this result into Eq.\ (\ref{ResponsFunct}) we find,%

\begin{equation}
\tilde{\alpha}_{jk}\left(  t\right)  =\frac{2\pi}{k_{\mathrm{B}}T}%
\Theta(t)\frac{d}{dt}\left\langle \delta p_{k}\left(  0\right)  \delta
p_{j}\left(  t\right)  \right\rangle . \label{ResponsDipole}%
\end{equation}

It is convenient to do a Fourier transform to express
Eq.\ (\ref{ResponsDipole}) in the frequency domain with $\tilde{\alpha}%
_{jk}\left(  t\right)  \rightarrow\alpha_{jk}\left(  \omega\right)  $ and
$\delta p\left(  t\right)  \rightarrow\delta\tilde{p}_{j}\left(
\omega\right)  $. The correlation function in frequency domain, $\left\langle
\delta\tilde{p}_{j}\left(  \omega\right)  \delta\tilde{p}_{k}^{\ast}\left(
\omega^{\prime}\right)  \right\rangle $ is%

\begin{align}
&  \left\langle \delta\tilde{p}_{j}\left(  \omega\right)  \delta\tilde{p}%
_{k}^{\ast}\left(  \omega^{\prime}\right)  \right\rangle \nonumber\\
&  =\delta\left(  \omega-\omega^{\prime}\right)  \frac{1}{2\pi}%
%TCIMACRO{\dint \limits_{-\infty}^{\infty}}%
%BeginExpansion
{\displaystyle\int\limits_{-\infty}^{\infty}}
%EndExpansion
\left\langle \delta\tilde{p}_{k}\left(  t\right)  \delta\tilde{p}_{j}^{\ast
}\left(  t+\tau\right)  \right\rangle e^{i\omega t}dt.
\end{align}
To obtain a spectral representation of the fluctuation-dissipation theorem, we
need to Fourier transform Eq.\ (\ref{ResponsDipole}). The right hand-side
leads to a convolution between the spectrum of the step function $\Theta(t)$
and the spectrum of $d/dt\left\langle \delta p_{k}\left(  0\right)  \delta
p_{j}\left(  t\right)  \right\rangle $. To remove of the imaginary part of the
step function, we solve for $\left[  \alpha_{jk}\left(  \omega\right)
-\alpha_{kj}^{\ast}\left(  \omega\right)  \right]  $ instead of $\alpha
_{jk}\left(  \omega\right)  $. Using the Wiener-Khintchine theorem\cite{MCJ09}
and taking the real part (note that $\left\langle \delta p_{k}\left(
t\right)  \delta p_{j}\left(  t+\tau\right)  \right\rangle $ is real) we get%

\begin{equation}
\left\langle \delta\tilde{p}_{j}\left(  \omega\right)  \delta\tilde{p}%
_{k}^{\ast}\left(  \omega^{\prime}\right)  \right\rangle =\frac{k_{B}T}{2\pi
i\omega}\left[  \alpha_{jk}\left(  \omega\right)  -\alpha_{kj}^{\ast}\left(
\omega\right)  \right]  \delta\left(  \omega-\omega^{\prime}\right)  .
\label{classical}%
\end{equation}
\qquad\ The Wiener-Khintchine theorem applies to classical systems, where
$\hbar\omega\ll k_{\mathrm{B}}T$. However, we can generalize it to quantum
systems by replacing $k_{B}T$ by $\hbar\omega/\left(  1-e^{-\hbar\omega
/k_{B}T}\right)  $ and substituting into Eq.\ (\ref{classical}). Then
$\left\langle \delta\tilde{p}_{j}\left(  \omega\right)  \delta\tilde{p}%
_{k}^{\ast}\left(  \omega^{\prime}\right)  \right\rangle $ becomes
\begin{align}
\left\langle \delta\tilde{p}_{j}\left(  \omega\right)  \delta\tilde{p}%
_{k}^{\ast}\left(  \omega^{\prime}\right)  \right\rangle  &  =\frac{1}{2\pi
i\omega}\left[  \frac{\hbar\omega}{1-e^{-\hbar\omega/k_{B}T}}\right]
\nonumber\\
&  \times\left[  \alpha_{jk}\left(  \omega\right)  -\alpha_{kj}^{\ast}\left(
\omega\right)  \right]  \delta\left(  \omega-\omega^{\prime}\right)  .
\label{FTquantum}%
\end{align}
Fluctuation is now converted to dissipation on the right hand side. We now
need to determine $\alpha_{jk}\left(  \omega\right)  -\alpha_{kj}^{\ast
}\left(  \omega\right)  $. Dissipation in the system is associated with the
imaginary part of the complex electric susceptibility, $\chi_{e}^{\prime
\prime}$, since $p\propto\varepsilon_{0}\chi_{e}$, where $\varepsilon_{0}$ is
the vacuum permittivity and $\chi_{e}$ is the complex electric susceptibility.
Therefore, the dissipation term is contained in $\chi_{e}^{\prime\prime}$, and
$\alpha_{jk}\left(  \omega\right)  \ $is thus $i\varepsilon_{0}\chi
_{e}^{\prime\prime}$.

The noise spectrum is obtained from the fluctuation of the dipole moment by
plugging $\alpha_{jk}\left(  \omega\right)  =i\varepsilon_{0}\chi_{e}%
^{\prime\prime}$\ in Eq.\ (\ref{FTquantum}) and then into
Eq.\ (\ref{SwithPcorrelation})%
\begin{equation}
S(\mathbf{k},\omega)=2\int_{-\infty}^{\infty}\left\langle \delta
p(\boldsymbol{r},\tau)\delta p(\boldsymbol{r}^{{\prime}},\tau+t)\right\rangle
e^{-i\omega t}dt.
\end{equation}
However, since the system's response function is local, fluctuations at
distinct spatial coordinates are uncorrelated. Then $\delta p(\boldsymbol{r}%
,\tau)=\delta p(\tau)$ and $S(\mathbf{k},\omega)=S\left(  \omega\right)  $, so
that
\begin{equation}
S\left(  \omega\right)  =\frac{4\hbar}{1-e^{-\hbar\omega/k_{B}T}}%
\varepsilon_{0}\chi_{e}^{\prime\prime}. \label{noise_final}%
\end{equation}

\end{document}